# Quenched QCD at finite density: $g = 1$ and $g = \infty$ [*] [†]

J. B. Kogut [a] M–P Lombardo [b] and D. K.Sinclair [c]

[a]Department of Physics, University of Illinois at Urbana–Champaign, 1110 West Green Street, Urbana, IL 61801-3080, U.S.A.

[b]HLRZ c/o KFA Jülich, D-52425 Jülich, Germany

[c]HEP Division, Argonne National Laboratory, 9700 South Cass Avenue, Argonne, IL 60439, U.S.A

We report on our ongoing effort to understand quenched lattice QCD at finite baryon number density. The quenched theory is sensitive to the baryon mass both at strong coupling and in the scaling region. However, we find that the quenched model is pathological for $\mu > m_\pi/2$ at $\beta = 6.0$, in agreement with past Lanczos analyses of the Dirac operator.

This note contains an upgrade of our work at $\beta = 6.0$ [1] to larger lattices, and a report of new results in the strong coupling limit. The considerations which motivated us to start a new program at finite density have been presented in detail in the past [1]. Our primary goal was to obtain accurate measurements of thermodynamic observables and spectra, which give complementary information on the physics of the transition. We have had partial success, since the results were satisfactory up to $\mu \simeq m_\pi/2$. Particularly interesting were the results for the baryon mass. Its behaviour was found to be in agreement with simple potential models challenged by previous lattice studies. A very modest bending in the chiral condensate was observed for $\mu \simeq m_\pi/2$. Unfortunately, in the interesting region, $\mu > m_\pi/2$, the fluctuations inherent in the simulation were so large that it was impossible to obtain sensible results. Then, as a further step, we decided to investigate these problems by moving to bigger lattices, and by using a different source for the inversion of the Dirac operator. We have thus obtained results at $\beta = 6.0$ that we shall present below.

We have also proposed an explanation for the problems afflicting the quenched model which relies on the fact that poles in the fermion propagator for $\mu \simeq m_\pi/2$ can indeed be there on isolated configurations, as observed in past work [1], [2], thus giving hints of deconfinement and chiral symmetry restoration. However, the contributions of these poles should cancel in the ensemble average of physical quantities, if confinement is realized. In our new work, we have searched for direct evidence of such poles by building an unphysical operator, a 'baryonic pion', and we have found the pathologies expected for $\mu > m_\pi/2$. These measurements were made at strong coupling, where, among other technical advantages, $m_\pi/2$ and $m_B/3$ are well separated, and the lattice data are also amenable to a cross check with analytic computations [3]. We begin now the exposition of our results.

At $\beta = 0$ we have generated 20 configurations on a $8^3 \times 16$ lattice. We have inverted the Dirac operator three times, for the three different boundary conditions [1], for each $m, \mu$. The bare quark mass was .1, $\mu$ ranged from 0. to 1.2, and the mean field prediction for the pseudocritical $\mu$ is $\mu \simeq 0.6$. We have measured the chiral condensate, the number density, the energy density and the pion mass. The other masses are too heavy, and difficult to extract. Along with standard observables, the unphysical operator $G_\mu G_\mu^\dagger$ (the 'funny' or 'baryonic' pion, as opposed to the real pion $G_\mu G_{-\mu}^\dagger$) was measured to obtain direct

[*]Partially supported by NSF under grant NSF-PHY92-00148 and by DOE contract W-31-109-ENG-38. Simulations done on the C90 CRAY's at PSC and NERSC
[†]Contribution to Lattice94, Bielefeld

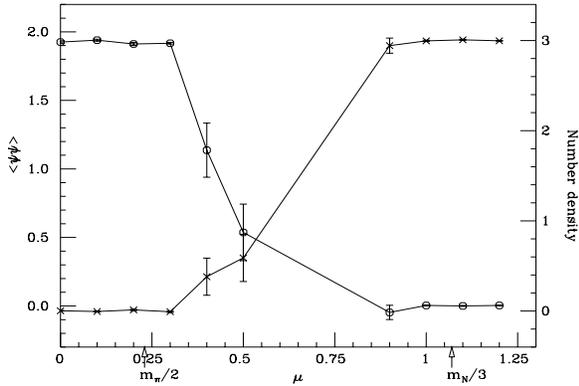

Figure 1. Chiral condensate (circles, left scale) and number density (crosses, right scale) as a function of $\mu$ at $\beta = 0$. The solid lines are to guide the eye.

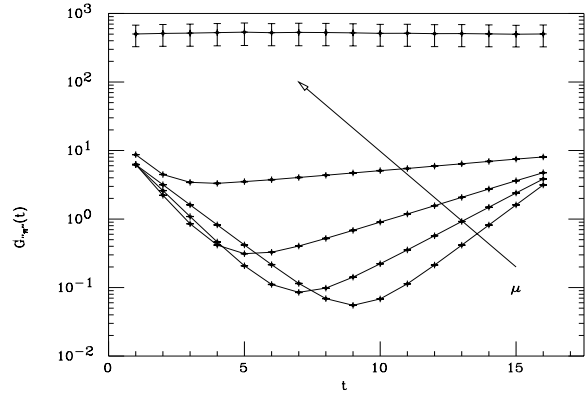

Figure 2. 'Funny pion' propagator at $\beta = 0$. $\mu = 0., 1, .2, .3, .4$, from top to bottom. Note the 'baryon–like' behaviour at small $\mu$, and the flatness for $\mu = .4$

evidence for poles of the Dirac operator.

We have observed a dramatic increase in the cpu time required for the inversion of the Dirac operator when entering the $\mu > m_\pi/2$ region. This is consistent with the electrostatic analogy [2], since in this region the Dirac matrix is almost singular. For $\mu > m_B/3$ the number of iterations needed for the inversion is again very small. In this region we are inside the eigenvalue 'crescent', and the first Brillouin zone is saturated: all the observables have their limiting values. The chiral condensate is zero and the number density is three (Fig.1). This tells us that the theory in the strong coupling limit 'knows' about the baryon mass.

We note that the average value of the chiral condensate is consistent with the one previously reported in the literature. However, for $\mu > m_\pi/2$ the results for $<\bar\psi\psi>$ obtained configuration by configuration spread over a wide range, which includes both zero and the $\mu = 0$ value. This suggests that we might have entered a metastability region rather than the symmetric phase, and that the identification of the onset for $<\bar\psi\psi>$ with the pseudocritical point computed in a mean field approximation may not be correct.

The behaviour of the 'funny' pion is very interesting. Its propagator has a baryon-like (i.e. T-asymmetric) behaviour for $\mu < m_\pi/2$, and then flattens 'inside' the eigenvalue cloud ($\mu > m_\pi/2$) (Fig. 2). We have clear evidence of zero modes in the quark propagator for $\mu \simeq m_\pi/2$, again in agreement with the eigenvalue picture. The mathematical and physical significance of this result is open to many interpretations which have been discussed and reviewed in the literature [1].

At $\beta = 6.0$ we have accumulated 30 independent configurations, on a $16^3 \times 64$ lattice. A subset of configurations was used for the inversion of the Dirac operator for each $\mu$ value. The quark mass was .02. The role of the winding loops was monitored by doubling (with respect to our previous work) the temporal extent of the lattice. Also, we decided to use a wall-noisy source to reduce the non–positivity effects associated with the rigid wall we were using before.

We confirm the decline observed in the chiral condensate (Fig.3). The huge and unphysical fluctuations in the pion propagator described in [1] disappear. However, the number of iterations required for the inversion when $\mu$ exceeds $m_\pi/2$ remains big, of order $10^4$. This means that the Dirac operator is still nearly singular. The source affects the amplitudes, but not the poles, as expected. Despite the better behaviour of the propagators, the mass estimates were not compelling: it is well known that a point source, even with the noisy improvement, has poorer behaviour than a rigid wall one. In particular, we were not able to measure the baryon mass.

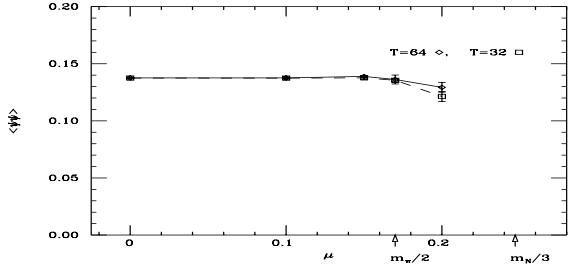

Figure 3. Chiral condensate as a function of $\mu$ on the $16^3 \times (32, 64)$ lattices, from a noisy estimator

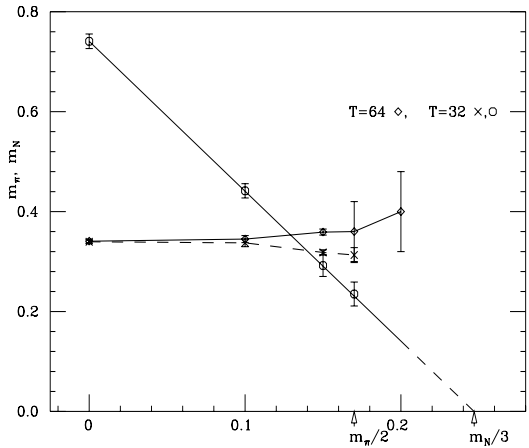

Figure 4. Pion and nucleon masses as a function of $\mu$ on the $16^3 \times (32, 64)$ lattices, from a (rigid, noisy) wall source.

We have some indication of an increase of the pion mass across the transition (Fig.4), consistent with the decline of the chiral condensate, and suggestive of chiral symmetry restoration. The effect is not dramatic ( The pion mass is basically constant within its statistical errors. Moreover, wall and noisy-wall sources give lower and upper bounds to the pion mass, respectively, so one may be tempted to 'average' the results, thus obtaining a constant pion mass), but it is corroborated by the trend in the propagators. In fact, we have observed that the pion and the sigma propagators become nearly degenerate.

Also in this case, as we did at $\beta = 0.$, we tried to get inside the eigenvalue distribution by increasing $\mu$ to $\mu = 1.0$. However, the number of iterations required for the inversion was still huge, meaning that we were still inside the eigenvalue cloud. In the free theory saturation occurs only for large $\mu$, and we are probably observing a similar trend at $\beta = 6.0$, consistent with the observations of [4]. So, the physics of the quenched model close to the continuum for $\mu > m_\pi/2$ still evades us.

In conclusion, is the finite density puzzle solved? Is there a phase transition at $\mu = m_B/3$, obscured by precursor phenomena? Or is the transition at $\mu = m_B/3$ really preempted by a spurious one at $m_\pi/2$? Clearly, something pathological is occuring at $\mu \simeq m_\pi/2$–it has prevented us from getting results for $\mu > m_\pi/2$ at $\beta = 6.0$. But it is also clear that the theory 'knows' about the baryon mass, and that it behaves in a sensible way at small values of the chemical potential. Our next step would be a full QCD simulation at finite temperature, and moderate values of chemical potential, to be run in parallel with the quenched model. We look forward to phenomenologically relevant results, and, by comparing the two models, to a better understanding of the failures of the quenched approximation. Of course, it is unclear whether simulation methods, designed for unquenched QCD, such as the complex Langevin algorithm, will work at nonzero $\mu$ since they depend on the existence of the quark propagator configuration by configuration.